\documentclass[letterpaper,twocolumn,10pt]{article}

\usepackage[letterpaper, hmargin=1in, vmargin=1in]{geometry}
\usepackage{times}
\usepackage{setspace}
\usepackage[nocompress]{cite}
\usepackage[hyphens]{url}
\usepackage{moreverb}
\usepackage{ifpdf}

\usepackage{multirow}

\ifpdf
\usepackage[pdftex,colorlinks, linkcolor={blue}, citecolor={blue}]{hyperref}
\usepackage{graphicx}
\else
\usepackage{graphicx}
\fi

\graphicspath{{fig/}}

\usepackage{amsmath}
\mathchardef\mhyphen="2D

\newcommand{\Unity}{Unity}
\newcommand{\Unityd}{Unityd}
\newcommand{\UBD}{UBD}
\newcommand{\UFS}{UFS}

\usepackage{epstopdf}

\begin{document}

\clubpenalty=10000 
\widowpenalty = 10000


\title{\Unity\ 2.0: Secure and Durable Personal Cloud Storage }
\date{}
\author{
{\rm Beom Heyn Kim, Wei Huang, Afshar Ganjali, David Lie}\\
University of Toronto}
\maketitle

\newcommand{\nmappingsfroyo}{17,218}
\newcommand{\nmappings}{29,208}
\newcommand{\napps}{1,260}


\renewcommand{\dbltopfraction}{1.0}
\renewcommand{\topfraction}{1.0}
\renewcommand{\bottomfraction}{1.0}
\renewcommand{\textfraction}{0.2}



\newcommand{\myfig}[5]
{
\begin{figure}[tb]
\begin{center}
\ifpdf
\includegraphics[width=#4\linewidth]{#1}
\else
\includegraphics[width=#4\linewidth]{#1}
\fi
\end{center}
\vspace{-20pt}
\caption{#2}\label{#3}
\vspace{#5}
\end{figure}
}

\newcommand{\myfigwide}[4]
{
\begin{figure*}[tb]
\begin{center}
\ifpdf
\includegraphics[width=#4\linewidth]{#1}
\else
\includegraphics[width=#4\linewidth]{#1}
\fi
\end{center}
\vspace{-20pt}
\caption{#2}\label{#3}
\end{figure*}
}

\newcommand{\BA}{{\em begin\_atomic}}
\newcommand{\EA}{{\em end\_atomic}}

\newcounter{claimcounter}[section]
\newtheorem{claim}{\sc Claim:}

\begin{abstract}
While personal cloud storage services such as Dropbox, OneDrive, Google Drive and iCloud 
have become very popular in recent years, these services offer few security guarantees to users.  These cloud services are aimed at end users, whose applications often assume a local file system storage, and thus require strongly consistent data.  In addition, users usually access these services using personal computers and portable devices such as phones and tablets, which are upload bandwidth constrained and in many cases battery powered.  \Unity\ is a system that provides confidentiality, integrity, durability and strong consistency while minimizing the upload bandwidth of its clients.  
We find that \Unity\ consumes minimal upload bandwidth for compute-heavy workload compared to NFS and Dropbox, while uses similar amount of upload bandwidth for write-heavy workload relative to NBD. Although read-heavy workload tends to consume more upload bandwidth with \Unity, it is no more than an eighth of the size of blocks replicated and there is much room for optimization. Moreover, \Unity\ provides flexibility to maintain multiple DEs to provide scalability for multiple devices to concurrently access the data with the minimal lease switch cost.
\end{abstract}


\section{Introduction}
A large number of personal cloud storage options, such as Dropbox, OneDrive, Google Drive, iCloud, Box, Sugarsync and SpiderOak exist today.  These services consist of an online storage server and automatic synchronization software that mirrors some or all of the files the user has stored in the cloud to the user's devices.  These services have millions of users who entrust their valuable data to these providers.  Unfortunately, the user must trust the cloud provider with the security of their data.  While users may encrypt data before uploading it to the cloud, this only guarantees confidentiality -- the integrity and durability of the data stored in the cloud is still in the hands of the cloud provider.  For example, the cloud provider could silently corrupt user data, and the corrupted files would be automatically replicated to all user devices replacing any trace of the correct data.  Even worse, a malicious or faulty cloud provider could delete or destroy files, causing the affected files to also be deleted on all the user's devices. 

In this paper, we describe the design and implementation of \Unity\ a system that protects the security and durability of data stored in the cloud.  While recent work~\cite{DBLP:conf/osdi/MahajanSLCADW10,DBLP:conf/osdi/FeldmanZFF10,DBLP:conf/ccs/ShraerCCKMS10} has addressed the problem for untrusted cloud storage.  Like previous systems, \Unity\ encrypts data before storing on the cloud to protect confidentiality and replicates blocks over the user's devices to ensure durability.  However, \Unity\ differs from previous solutions in two ways.  First, the previous proposals do not protect against an adversary that weakens the consistency of data stored in the cloud and instead require on the application layer to deal with weakly consistent data.  Because \Unity\ is designed for personal cloud storage, we assume all client devices are used by a single user, and thus design \Unity\ around the notion that concurrent access by different devices to the same object is rare and does not have to be fast.  Moreover, many of the end-user applications users use to access files stored in a personal cloud were designed around a local file system and thus may require some level of consistency.  As a result, in designing \Unity, we opted to enforce strong consistency in the face of failures and even against a malicious adversary~\cite{SalusNSDI13}.  \Unity\ enforces consistency at the granularity of a storage abstraction we call a {\em data entity} (DE).   DEs can be mapped to various application-level storage objects, such as files or an entire file system, thus allowing the system designer to pick different points in the trade-off between performance, scalability and consistency.

Second, users tend to access personal cloud services from devices that have upload bandwidth-constrained network connections.  To detect server equivocation, individual devices must communicate directly with each other to prevent the server from hiding the writes of one device from another~\cite{sundr}.  However, excessive communication between devices depletes this limited upload bandwidth.  For example, the authors of Depot report that gossiping to detect server equivocation increases the bandwidth requirements of writes by approximately 50\%~\cite{DBLP:conf/osdi/MahajanSLCADW10}.  \Unity's strong consistency allows it to minimize inter-device communication because the order of writes across devices is well-defined.  As a result, devices need only directly communicate when the writer of a DE changes or to check whether a device has failed or not.

Our main result is demonstrating that an alternative ``contract'' is viable between an individual user and the cloud storage provider.  The standard relationship has had the user pay the cloud provider for the durability and security of their data.  However, we observe that the most difficult properties for an end user to achieve are availability and high upload bandwidth for their devices.  User devices are on networks that may fail intermittently or the devices themselves may crash or run out of battery.  Similarly, high upload bandwidth connections are expensive and can rapidly consume a device's limited battery.  On the other hand, users have increasing numbers of personal devices and those devices have had increasingly larger and larger local storage.  As a result, it is becoming easier for the user to replicate and provide durability for data themselves.  \Unity\ proposes a new relationship where security and durability are provided by the user and the cloud provider provides availability of the data and network bandwidth to quickly replicate the data.

\vspace{2pt}
\noindent This paper makes the following contributions:
\begin{enumerate}
\item We describe the design and implementation of \Unity\ a system that uses a cloud provider's availability and network bandwidth to efficiently provide security and durability for personal data stored in the cloud.
\item We compare the performance and bandwidth utilization of \Unity\ against centralized server solutions such as NFS and NBD, as well as existing cloud solutions such as Dropbox. 
\Unity\ could perform comparably to NFS, NBD and Dropbox. Also, we shows potentially there can be large upload bandwidth consumption saving over various networks using \Unity. 
\item We study how the DE abstraction can be mapped to a coarse-grain storage unit such as an entire file system or to fine-grain storage units such as individual files on a modified version of the Minix file system.  We also measure the trade-off between the improved performance of fine-grain DE mappings due to less false sharing and the higher book-keeping costs and overhead introduced by the management of the larger number of mappings.
\end{enumerate}

We present the architecture overview of \Unity\ in the Section~\ref{sec:overview}. In the Section~\ref{sec:implementation}, core data structures, \Unity\ interface and the various component of our current \Unity\ prototype are described in detail. Evaluation results are described in the Section~\ref{sec:eval} followed by the related work summary in the Section~\ref{sec:related}. At last, we conclude the paper with the section~\ref{sec:conclusion}.

\section{Overview}
\label{sec:overview}
We begin by describing the basic components of \Unity\ and give a simplified description of the protocol it uses.  We then enhance this protocol to provide durability, enforce security and deal with failures.

\begin{table*}
\begin{center}
\begin{tabular}{|l|l|p{0.65\linewidth}|}
\hline
\multicolumn{1}{|c|}{\bf Message} &
\multicolumn{1}{|c|}{\bf Participants} &
\multicolumn{1}{|c|}{\bf Description} \\ \hline \hline
State fetch & Device$\rightarrow$CR & Device requests all state updates for a DE starting at a particular sequence number. \\ \hline
Block request & Device$\rightarrow$CN& Device requests contents of a block-version. Cloud node will respond with appropriate block contents. \\ \hline
Write update & LH$\rightarrow$CR\&CN & Lease-holder sends a new block-version to the coordinator and cloud node.  The update must signed and include a sequence number.  Block contents must be encrypted. \\ \hline
LH-switch & Device$\rightarrow$CR & Device requests the lease for a DE. \\ \hline
LH-revocation & CR$\rightarrow$LH & Coordinator requests the lease back from the lease-holder. \\ \hline
LH-update & LH$\rightarrow$CR & Lease-holder gives up the lease and writes a record to the coordinator indicating the new lease-holder. The update must be signed and include a sequence number.  \\ \hline
LH-transfer & LH$\rightarrow$New-LH & Current lease-holder transfers the lease to the new lease-holder.  The transfer must be signed and include the latest sequence number. \\ \hline
Replication update & Device$\rightarrow$CR & A device has replicated a block-version and notifies the coordinator. The update must be signed. \\ \hline
\end{tabular}
\end{center}
\caption{Major protocol messages in \Unity.  In the participants in a message can be the lease-holder (LH), the cloud node (CN), the coordinator (CR) or a non-lease-holder user device (Device).}
\label{tbl:message}
\end{table*}

\subsection{Basic system architecture}
\label{sec:basic}
A \Unity\ cloud consists of several user-controlled devices and a cloud storage provider who provides the storage service.  The cloud provider itself consists of two sub-components -- a {\em coordinator} and a cloud storage node, which we will refer to simply as the {\em cloud node} for brevity.  All user devices send regular heartbeat messages to the coordinator so that the coordinator can detect if a device has failed or lost network connection. 

Data stored in the cloud consists of fixed-sized blocks of data and are grouped into units called {\em Data Entities} (DE), which are the basic unit of consistency in \Unity.  Only one user device may read or write blocks stored in each DE at any given time.  To do so, the device must first obtain a lease from the coordinator.   Once obtained, a correctly working user device does not actively give up a lease on a DE -- it only give it up when another user device requests a lease for the same DE or it may lose the lease if it fails to send a heartbeat.  Our current implementation of \Unity\ provides strong consistency by requiring the exclusive lease for both reads and writes.  By only requiring a lease for writing instead of both writing and reading, \Unity\ can also provide sequential consistency for better performance.

Each DE is an array of blocks and each data block has a unique index within the DE.  In addition, writes to blocks in the DE are versioned and the cloud node and user devices may store various block-versions from DEs.  A correctly functioning cloud node will maintain storage for every block-version of every DE necessary to create a consistent image of each DE, while user devices may each store subsets of these block-versions to enable recovery against a malicious or failed cloud provider.  User devices can request a block stored on the cloud node by sending a {\em block request} that specifies the DE, block index and version of the desired block.

When the lease-holder for a DE writes a block, it creates a {\em write update} message, which contains the DE, block index, a version number and a hash of the block contents.  The lease-holder buffers write updates and the new block contents and periodically flushes the write updates to the coordinator and the block contents to the cloud node.  The coordinator records all such write updates in an append-only log for each DE, which we call the {\em state} of the DE.  Devices can request a copy of this log at any time from the coordinator using a {\em state fetch} request.  Both the cloud node and user devices store block contents in a key value store that is indexed by DE, block and version number.

When another devices wishes to access a DE, they send a {\em lease-holder switch} request to the coordinator, which then sends a {\em lease-holder revocation} message to the current leaseholder.  The current lease-holder sends any pending updates to the directory and then appends a {\em lease-holder update} message to the log, which contains the identity of the new leaseholder.  The current lease-holder then sends a {\em lease-holder transfer} message directly to the new lease-holder to indicate that it has finished flushing all the state to the coordinator and is now ready to give up the lease.  Before taking on the lease, the new lease-holder will request the state of the DE from the coordinator and thus become aware of the latest versions of all blocks.

With the wide availability of cellular data networks, we expect user devices to be connected to the network most of the time.  However, in the event that a lease-holder is not connected, \Unity\ provides a form of disconnected operation similar to Coda~\cite{DBLP:conf/sosp/KistlerS91}.  If the lease-holder misses a heartbeat message, then the coordinator labels the lease-holder as disconnected.  The remaining devices will run a lease-holder recovery protocol and a new lease-holder will be selected (described in more detail Section~\ref{sec:failures}.  If the disconnected device has outstanding writes when it reconnects to the cloud, it will have to check the DE state on the coordinator if any other writes occurred while it was disconnected.  If there were, it will have to resolve the conflicting writes.  Otherwise, it can simply make a lease-holder switch request, acquire the lease and send its outstanding writes to the cloud provider.

\subsection{Providing durability}
\label{sec:durability}

The system described above allows each user device to access data stored on the cloud, but does not provide any more guarantees than current commercial cloud providers.  To provide durability in the event of a cloud provider failure, \Unity\ replicates user data cross the user's devices up to some pre-determined {\em replication target}.  Furthermore, \Unity\ guarantees that each DE can be replicated to a consistent snapshot of its contents, similar to the prefix semantics guaranteed by Salus~\cite{SalusNSDI13}.  To provide this guarantee, \Unity\ must replicate in the order they were written and must also replicate older versions of blocks even if a newer version exists.  

We enhance each write update in the DE state with a list of devices that have replicated the block.  Initially, a newly written block will only have been replicated twice, once by the lease-holder that wrote the block and once by the cloud node.  If the replication target is greater than two, then other devices must also replicate the block to achieve the desired level of durability.  Other devices can find out which blocks require replication by requesting the state from the coordinator.  These devices can then request blocks by sending block requests to the cloud node.   After a device replicates a block, it updates the coordinator by sending a {\em replication update} that specifies the DE, block index and version that the device replicated, and the coordinator than appends the ID of the device to the corresponding write update in the DEs state.  A simpler implementation would only keep track of the number of replicas for each write update, not the actual devices storing the replicas, but \Unity\ does this to protect against a malicious coordinator as discussed in Section~\ref{sec:security}.  

A na\"{i}ve implementation would have each non-lease-holder device try to replicate the oldest block in the state of a DE.  However, because the devices update the DE state after they have replicated the blocks, this can result in races that cause blocks to become over-replicated.  Instead, for each DE, non-lease-holder devices locally compute non-overlapping subsets of write updates to replicate by hashing the outstanding blocks into a hash space of all non-lease-holder device IDs and replicating the blocks that collide with their device ID.  An alternative would have been to perform random block replication as in RAMCloud~\cite{ramcloud}, but we found that \Unity\ derived very little benefit from this approach.

Requiring the cloud node and user devices to retain old versions of blocks indefinitely would eventually consume all the storage on these machines.  As a result, nodes in \Unity\ garbage collect blocks if there is a more recent block that reached the replication target and that can be used to recover user data to a consistent snapshot from a device or cloud failure.  Determining which blocks can be garbage collected can be done locally by each device and the cloud node as long as they have up-to-date information about the {\em replication level} of each block-version, which is basically the number of currently existing replicas and can be obtained from the coordinator.  A malicious coordinator could reduce durability by lying about which the replication level, causing blocks to be prematurely garbage collected.  We will show how \Unity\ prevents this attack below.

\subsection{Providing Security}
\label{sec:security}

\Unity\ delivers confidentiality, integrity and durability for user data despite a malicious cloud provider.  \Unity\ defines a malicious cloud provider as a coordinator and/or cloud node that does not follow the above protocol.  \Unity\ does not differentiate between the reasons for cloud provider misbehavior, which could be due to malicious intent, operator error or software bugs. 

User devices can also act maliciously due to a malware infection or theft by a malicious individual.  \Unity\ also protects against malicious user devices but to a lesser extent.  \Unity\ cannot protect the confidentiality or integrity of user data since all user devices have privileges to read and write the user's data.  For example, \Unity\ does not prevent a malicious user device from deleting or overwriting the user's data.  However, \Unity\ does ensure that a malicious user device cannot subvert the durability guarantees against a malicious cloud provider.  \Unity\ limits a user device from reducing the replication level of any block by at most one (the malicious user device itself).  In this way, malicious user devices can do as much harm as a failed device with respect to durability. 

\Unity's guarantees are predicated on three assumptions.  First, a malicious user device and a malicious cloud provider do not collude.  If they do, than the malicious cloud provider will gain the ability to subvert the confidentiality and integrity of the user's data via the malicious device, breaking the set of guarantees that \Unity\ places against a malicious cloud provider.  Second, we assume that devices have reasonably accurate local clocks.  This is necessary because as described below, the lease-holder needs to know if it has missed a heartbeat due to a device failure or a network failure.  Third, we assume a network failure model where a device is either able to communicate with all other devices and the cloud provider or not able to communicate with any of them.  This assumption simplifies failure detection of devices.

We secure \Unity\ by further enhancing the protocol to deal with attacks against the confidentiality of data, the integrity of data and the durability of data.  

\noindent{\bf Confidentiality:} All user devices have a shared key that they use to encrypt and decrypt data stored on the cloud node.  The cloud provider is not aware of this key and thus cannot read any of the data stored on the cloud node, thus guaranteeing confidentiality.  Our current implementation of \Unity\ does not allow the cloud provider to de-duplicate identical blocks across users.  However, \Unity\ can also use convergent encryption~\cite{farsite,DBLP:conf/storagess/StorerGLM08}, which allows de-duplication at the cost of some amount of privacy for users.

\noindent{\bf Integrity of block contents:} A malicious cloud provider may attempt to modify the block contents stored on the cloud node after it is uploaded by the lease-holder.  To prevent this, each user device has a public-private key pair for signing where the public key half is known to all other user devices.  The lease-holder signs each write update that contains a hash of the block contents for the corresponding block version.  All devices verify the signatures of write updates received from the coordinator and verify the contents of blocks received from the cloud node using the hash in the corresponding write update.

\noindent{\bf Integrity of DE state:} Since all write updates are signed, a malicious coordinator cannot tamper with write updates or forge new ones.  However, there are still two attacks it may perform.  First, it may mount an omission attack by dropping write updates from the DE state.  Second, as mentioned earlier, it may tamper with or forge new replication updates to artificially increase the apparent replication level of a block.  This attack decreases durability by causing the tampered block to get under-replicated and can cause other blocks to get prematurely garbage collected.

We enhance each write update with a sequence number, which must be unique for each DE.  Each write update and lease-holder update message will have a unique, monotonically increasing sequence number and the sequence number shall be large enough so that it is unlikely to wrap (64 bits in our implementation).  This not only prevents the omission of a write update or lease-holder update message, but also puts a well-defined order on these events that is determined by the lease-holder that writes the messages into the DE state.  The lease-holder is required to write an update into the state at a regular heartbeat interval even if it has no new blocks to write, allowing devices to detect if a malicious coordinator truncates the log because it will not see an update for a period that exceeds the heartbeat interval.  When lease-holders transfer the lease to another device, they flush any messages they have to the coordinator. Then, they include the signed sequence number of the last message sent to the coordinator in the lease-holder transfer message that is sent directly to the new lease-holder.  This short message prevents the coordinator from showing the lease-holder an incomplete view of the DE state and constitutes the only inter-device communication that occurs during the regular operation of \Unity.

To prevent the coordinator from tampering with replication updates, each user device that produces the replication update must sign the update.  This prevents the coordinator from forging new replication updates, which would cause blocks to get under-replicated or other blocks to get garbage collected.  \Unity\ does not prevent omission attacks on replication updates as there is not strict ordering that can be imposed on these updates.  Since replication is asynchronous, even if a sequence number were used, there could be duplicate replication updates with the same sequence number under benign conditions due to races.  If the coordinator mounts an omission attack on the replication update, all she would do is causing the block to get over-replicated.  While this wastes both storage and download bandwidth for user devices, this does not affect the durability of the user's storage. We do not consider this attack undesirable, because it does not violate the guarantees \Unity\ seeks to provide.

\noindent{\bf Summary:} \Unity\ protects against tampering and information leakage by a malicious cloud provider using a combination of encryption to protect confidentiality, and digital signatures and sequence numbers  to protect the integrity of the block data and DE state.  We note that unlike typical personal cloud services, \Unity\ does not support sharing.  While \Unity\ could be extended to allow read sharing by providing other users with the encryption key, or read and write sharing by also providing them with a signing key, allowing multiple users to access the same DE may break the assumption of low concurrency.  Instead, we see \Unity\ as complementary to cloud services that support sharing.  A user can use \Unity\ to store sensitive data as well as data they want to be highly durable, and use a conventional cloud service to store data that they want to share.  

\subsection{Dealing with failures}
\label{sec:failures}

There are roughly two ways components in the \Unity\ cloud can fail.  They can become unavailable, and either return later or never return, or they can become malicious, meaning that they act dishonestly with respect to the \Unity\ protocol.  \Unity\ deals with both types of failures for both the cloud provider and user devices.

\noindent{\bf Cloud provider failure:} Temporary unavailability of the cloud provider is dealt with using a standard time-out after which the temporary unavailability is treated as permanent unavailability.  Similarly, if a cloud provider is found to be malicious because a signature check fails or it has omitted updates, the user devices will stop using the cloud provider and treat it as failed.  When this happens, the user of the devices needs to associate them with a new cloud provider.  The devices initialize the new cloud provider by combining the locally cached DE states and then can populate the cloud node with their replicated cloud blocks.  In this case, no data is lost since the contents of every block will have been replicated and the lease-holder for each DE will have an up-to-date state for that DE.

\noindent{\bf Unavailable user device:}  There are two cases for user device unavailability -- when the user device is not a lease-holder for some DE and when the user device is a lease-holder.  The coordinator will detect device failure if the device fails to send a heartbeat message or if the device fails to respond to a lease-holder revocation message.  If the device is not a lease-holder, the failure is straightforward to deal with.  The other devices can verify this by trying to contact the failed device directly.  If this fails, the device's replication updates are removed, prompting other devices may have to start re-replicating those blocks.  When the device returns, it can resend its replication updates to the coordinator if it still has the blocks.  This is why \Unity\ maintains a directory of which nodes have replicated which blocks instead of just maintaining a count.

\noindent{\bf Unavailable lease-holder:} If the unavailable device is a lease-holder, then \Unity\ must select a new lease-holder for the DE.  However, it is absolutely critical in selecting a new lease-holder that one and only one lease-holder take the place of the failed lease-holder.  The reason is that having 2 leaseholders can allow a malicious coordinator to obtain two different signed write updates with the same sequence number, allowing it to fork the state for the DE.  Ensuring this guarantee can be broken into two sub-problems.  First, the devices must ensure that the failed device has really failed and that a malicious coordinator is not claiming a failed device when the device is still available.  Not doing this check would cause another node to become the lease-holder when the original lease-holder still thinks it has the lease.  The devices can protect themselves against this attack by directly contacting the current lease-holder to ensure it really has failed.  If an unavailable lease-holder returns, perhaps after recharging its battery or reconnecting to the network, it will realize that it has missed a heartbeat due to having an accurate local clock, and return as a non-lease-holder.  It will then have to reacquire the lease before uploading any outstanding writes that it as.

Second, once the devices determine the current lease-holder has failed, they must select a new one and all agree on the new one.  While this can be done with a distributed consensus protocol like Paxos~\cite{paxos}, \Unity\ offers a much simpler way to do this.  Up to this point, \Unity's protocol ensures that all devices have common prefix of the DE state and the protocol ensures that the coordinator cannot indefinitely truncate the DE state from devices.  Thus, all devices will eventually see the exact same lease-holder update message in the state from the previous lease-holder.  Thus, in \Unity, once devices have determined the current lease-holder has failed, they all switch to using the previous lease-holder as the current lease-holder.  If it turns out that has failed, they can continue going backwards in the state to find the previous lease-holder to that until the find that some previous lease-holder is already one of the ones they have tried.  At this point, recovery of the DE will require human intervention by the user.  

\noindent{\bf Malicious user device:} As mentioned earlier,  \Unity\ cannot protect the confidentiality or integrity of user data from a malicious device, but limits the impact a malicious device has on durability to be the same as an unavailable device.  In addition, a malicious device can gain the lease, but must give it up if requested since failure to respond to a lease-holder revocation message will get the device labeled as failed.  This does not necessarily implicate the device because other devices cannot tell the difference between a malicious user device and a malicious cloud server, it will alert the user that something is amiss.  Finally, a malicious device may attempt to fork the state of a DE by creating several copies of a block-version with the same sequence number.  However, since the coordinator is responsible for distributing the write updates to the other user device, it needs the cooperation of the coordinator to carry out the fork attack.  This is why \Unity's guarantees only hold if a malicious user device does not collude with a malicious cloud service.

\begin{figure}[!htb]
  \centering
  \includegraphics[scale=.3]{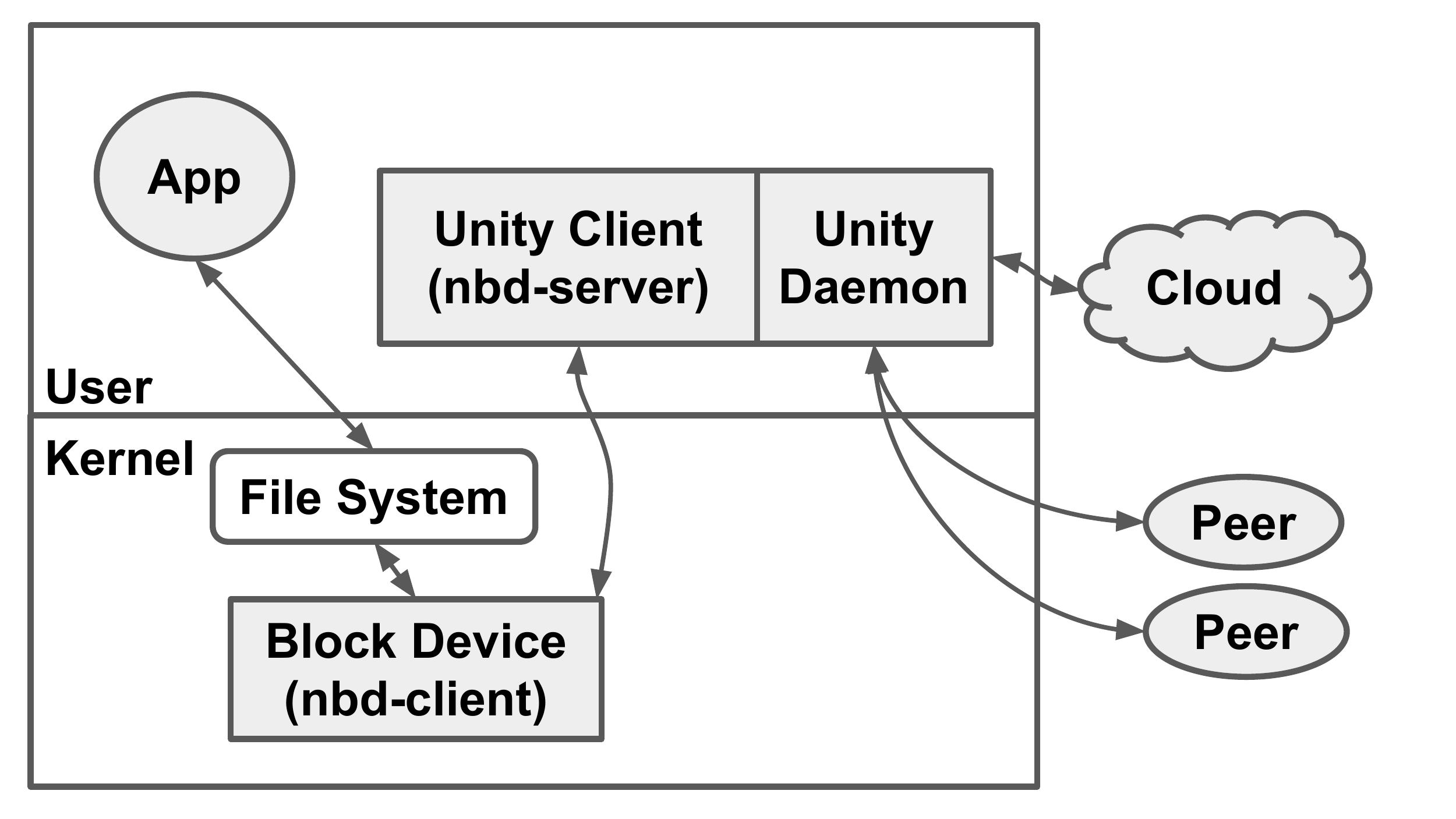}
  \caption{Client Node Architecture}
  \label{fig:Unity_Architecture}
\end{figure}

\begin{table*}
\begin{center}
\begin{tabular}{|p{0.30\linewidth}|p{0.65\linewidth}|}
\hline
\multicolumn{1}{|c|}{\bf API Call Prototype} &
\multicolumn{1}{|c|}{\bf Description} \\ \hline \hline
init (ul\_config\_t conf) & Initialization function taking various parameters.  Called before starting \Unityd.  Client may choose to create DEs after initialization and before starting \Unityd.  \\ \hline
cleanup (void) & Clean up function before disconnecting device from the \Unity\ cloud. \\ \hline
controller\_start (pthread\_t *thread) & Starts up \Unityd. \\ \hline
controller\_stop (pthread\_t thread) & Stops the \Unityd. Should be used with clean to terminate\\ \hline
create\_entity (u\_int64\_t  ID, u\_int64\_t entity\_size) & Used to create a new DE of specified size and ID.  The client must ensure that it is called with a unique ID every time. \\ \hline
read (u\_int64\_t ID, u\_int64\_t blockID, int offset, char *buf, int size) & 
Reads data of specified size into buffer from offset within a particular block.  Will always read the latest version block for a particular block ID.  \\ \hline
write (u\_int64\_5 ID, u\_int64\_t blockID, int offset, char *buf, int size) & 
Writes data of specified size from buffer to offset within a particular block.  This will create a new version of the block. \\ \hline\hline
callback\_revoke\_lease (u\_int64\_t ID) & Callback function used to notify client to prepare for the lease-holder switch for the DE specified by ID. \\ \hline
\end{tabular}
\end{center}
\caption{API calls in \Unity\ Library.  There are two types of API calls, library calls made by the client and callback functions called by \Unityd, which the client must supply.}
\label{tbl:api}
\end{table*}

\section{Implementation}
\label{sec:implementation}
Our \Unity\ prototype is implemented as in two components, a standalone coordinator, which runs as a simple event-driven server, and a \Unity\ daemon, called {\em \Unityd}, which runs on user devices and the cloud node.  \Unityd\ can be specialized by the type of a {\em client} linked against a library component of \Unityd.  We have implemented three types of clients: (a) \Unity\ Block Device (\UBD), which emulates a Linux block device, on top of which one can mount any standard file system; (b) \Unity\ File system (\UFS), which implements a Linux file syste`m; and (c) a cloud node client, which is \Unityd\ adapted to a cloud node.  The coordinator consists of about 4K lines of C code and \Unityd\ consists of about 15K lines of C.  Our clients vary between 2K-3.6K lines of C code.  As an example, Figure~\ref{fig:Unity_Architecture} illustrates an example of a \Unity\ device with a \Unity\ client that implements a simple block device for a generic file system to use.

%

\subsection{\Unityd}

\Unityd\ consists of four major components: (a) the library of API calls and callbacks that is linked with the client; (b) the state buffer, which buffers uncommitted DE state messages to the coordinator; (c) the block store, which stores local copies of block contents; and (d) three threads, which handle various events and operations in \Unityd.   

As mentioned, \Unityd\ contains three threads such as a controller thread, a client thread and a replication thread. A replication thread replicates blocks for durability; a client thread handles client requests such as read or write operations and lease-holder switching; and a controller thread handles block requests from other devices and updates to the coordinator. For more detailed explanation about them, we refer readers to the `Thread' paragraph below in this section.


\noindent{\bf Library interface:}  The \Unityd\ library exposes a simple API to the client consisting of 7 library calls and 1 callback function that the client must provide as described in Table~\ref{tbl:api}.  The first 4 API calls are only used during starting and shutdown of the device and the remaining 3 calls are used for creating DEs, reading from DEs and writing to DEs.  The two callback functions are used by \Unityd\ to notify the client of lease-holder switch events.  The {\tt callback\_revoke\_lease} callback is called when the client is about to lose the lease.  In general, this should cause the client to flush any writes it is caching to \Unityd\ so that they can be sent to the cloud provider.  The client should complete these flushes before returning from the callback function.  If the client calls {\tt read} or {\tt write} on a DE it doesn't have the lease for, \Unityd\ library will automatically acquire the lease before it returns.

The client itself is responsible for assigning a unique ID to each DE when it is created.  This is because it is unsafe for clients to trust the coordinator to do this since malicious coordinator can return the same DE ID for requests from two different devices in order to full clients and obtain two different signed updates for the same block.  To ensure that clients always pick a globally unique DE ID, we leverage a DE itself; more specifically, the special {\em bootstrap} DE (DE ID zero in our prototype) is used to maintain the largest used DE ID.  When allocating a new DE, clients atomically read and increment the value stored in the bootstrap DE and use this value as the ID for the new DE.  Since \Unity\ guarantees strong consistency, all clients are always guaranteed to see the consistent contents for the bootstrap DE.

\noindent{\bf State buffer:}  The {\em state buffer} stores a local copy for the state of every DE.  It also buffers new {\em write updates}, {\em lease-holder updates} and {\em replication updates} generated by the device that have not been sent to the coordinator yet.  To store this information, an in-memory data structure is implemented to store write updates and lease-holder updates as a per-DE append-only log format.  Each write update contains the block ID, the version number for the block, the sequence number for the update, the array of device IDs replicating the block, the block content's SHA1 hash and a 2048-bit RSA signature of the entire update.  Any replication updates are stored with the associate write update along with their respective signatures as well. Lease-holder updates store the sequence number and the IDs of the old and new lease-holders, along with a RSA digital signature.  

The state buffer maintains two indexes for easy searching.  A {\em block-version index} enables lookup by a \{block, version\} tuple and a {\em sequence number index} enables lookup by a sequence number.  The block-version index is used to obtain the hash of a block content to verify the integrity of the block content when it is fetched from the cloud node. For efficiency, rather than sending write or replication updates to the coordinator right after they are created, \Unityd\ will batch them and send them at a fixed interval (usually 30 or 60 seconds).  The sequence number index is used to easily find all the updates that were created since the last batch and needed to be sent to the coordinator.

The in-memory append-only log for each DE is accessed by either the client thread or replication thread which adds new updates to the state. Also, a controller thread accesses it to send new updates to the coordinator. Thus, there is race between them. In order to minimize contention, the state buffer uses a {\em double buffering} scheme to allow threads to concurrently access it without having to acquire locks.

\noindent{\bf Block store:}
The {\em block store} stores the block contents that make up the DEs.  It stores block contents downloaded from the cloud node including old block-versions to guarantee durability. Also, it stores current block versions to service the read requests received from the local client.  The block store in our prototype is implemented using the combination of a LevelDB key-value store and a large linear buffer on an ext4 file system. The key used for LevelDB is the concatenation of the 64-bit DE ID, 64-bit block ID and then 32-bit version number. Each key is mapped to the value that is the offset into the buffer where the actual contents of the corresponding block-version are stored in the file on the ext4 file system.

\noindent{\bf Threads:}
The replication thread reads the state buffer and assembles a list of block-versions that require replication for each DE.  Determining blocks to replicate is depending on whether the replication thread is running on a user device or on the cloud node.  If it is on a user device, it will only replicate the subset of the blocks that require replication.  As explained in Section~\ref{sec:overview}, the replication thread hashes under-replicated blocks into a hash space consisting of the device IDs of all non-lease-holder devices and begins replicating the blocks that collide with its device ID.  At the every heartbeat period, the new set of updates is fetched from the coordinator, triggering the replication thread to re-evaluate the set of blocks it needs to replicate.

On the cloud node, the replication thread will always replicate from the lease-holder and replicates every block that requires replication.  Thus, our implementation differ slightly from the architecture described in Section~\ref{sec:overview} since new block contents produced by writes by the lease-holder are actually fetched from the lease-holder by the cloud node as opposed to being pushed to the cloud node by the lease-holder.  This small change was done just to allow code re-use between user devices and the cloud node.  

The client thread implements the functions in the library interface and is not actually a stand-alone thread.  Instead, it runs in the same thread-context as the client.  When the client calls {\tt read} or {\tt write}, it will check if the device currently holds the lease for the DE.  Otherwise, it requests the lease from the coordinator and blocks the client until the lease is transferred.  Once it has the lease, it signals the controller thread to fetch the latest state information for the DE from the coordinator and then looks up the latest version for the block being requested.  Also, it checks if the content of the block-version is in the block store and if not, it makes a request to the cloud node to retrieve the block content.  Once it receives the content, it then verifies its integrity by computing a hash over the content and comparing that with the hash in the state.  

The client thread will access the state of a DE only if the device holds the lease for that DE, while the replication thread will access the state of a DE only if the device does not hold the lease for the DE.  Thus, the accesses to the state of a DE by the client and the replication thread are automatically mutually exclusive.

The controller thread has three functions.  The main function of the controller thread is fetching state updates from the coordinator and sending batches of updates from the state buffer to the coordinator.  The frequency at which it does this is equivalent to the heartbeat period of the device, which is 30 seconds for non-battery powered devices and 60 seconds for battery powered devices.  The cloud node sends the heartbeat at every 10 second to minimize the delay for it to learn about new updates.  Computing the signatures for updates to send to the cloud is deferred until the time that the batch needs to get sent over. Meanwhile, the signatures are stripped off once they are verified after being fetched from the coordinator.  Thus, no signature is stored in the state buffer of a device.  To reduce the bandwidth consumption, the controller tries to fetch only updates for blocks that have not been replicated as many times as the replication target.  Thus, it remembers the sequence number of the oldest block that has not been fully replicated and sends this to the coordinator when fetching updates.  The coordinator then returns all blocks with sequence numbers later than the specified sequence number.

The second function of the controller thread is to handle lease-holder switch requests.  The lease request for a DE sent to the coordinator can be rejected (NACKed) if the coordinator is already in the middle of a lease-holder switch for the same DE.  Then, the controller will back off and retry after some time interval has passed.  

The final function of the controller thread is handling block fetch requests from other devices.  If the controller is running on a user device, it will get requests from the cloud node and therefore upload block updates to the cloud. On the other hand, if the controller is running on the cloud node, it will offer the block to the lease-holder and handle the replication requests from other user devices.  

In all devices, the controller thread has the lower priority than the client thread.  The consequence of this is that the client thread's performance is preferred to that of the controller thread. Thus, if there is a lot of client I/O, replication requests from the cloud node can be delayed as the controller thread may be delayed to run.  However, this is acceptable as it actually leads to better I/O performance for applications running on the device that the user is currently actively using.

\subsection{Coordinator}

The coordinator is a simple, single-threaded server that services requests to fetch and update state information for DEs.  It uses essentially the same code used to manage the state buffer in \Unityd\ except that it is unable to sign or verify any of the signatures.  As a result, a slight modification is made to allow it to store the signatures along with the updates.  Devices fetch all updates for a DE greater than some sequence number.  To service these requests quickly, the coordinator uses the sequence number index to find the update from which it can start collecting updates to put into the response. 

The other function of the coordinator is to detect device failures and to inform the other devices.  So, it maintains a table for all connected devices, which contains their device ID, IP address, their heartbeat period and time passed since the coordinator lastly received a message from those devices. This table can also be fetched by user devices.

\subsection{Clients}
We implement two different \Unity\ clients to demonstrate the flexibility of \Unity\ prototype. The first is a \Unity\ Block Device (\UBD) which maps a block device onto a single large DE.  Devices can mount local file systems on \UBD.  However, if two devices try to concurrently access different files on the same file system, they will need to exchange leases, which hurts the performance.  To alleviate this false sharing, we demonstrate the flexibility of DEs using our second client, the \Unity\ File System (\UFS) where each file in a file system is mapped to its own DE.  Finally, a third client is used to implement the cloud node.  


\noindent{\bf \Unity\ Block Device:}
\Unityd\ is a user-space daemon but clients must in general have kernel component to allow \Unity\ to flush the kernel's buffer cache before a lease-holder switch.  We thus build \UBD\ on top of Network Block Device (NBD) version 2.9.25.  NBD consists of an nbd-client, which is a kernel-level block device and an nbd-server, which is a user-space server.  We run nbd-client and nbd-server on the same machine and use nbd-client to capture block requests from a local file system and send them to the user-space nbd-server.  We then modify nbd-server to translate block requests into library calls into \Unityd's client library.  The nbd-server can receive variable length requests, which \UBD\ pads or partitions as needed into 4KB block-sized requests to \Unityd.  The {\tt callback\_revoke\_lease} callback uses internal kernel functions to flush the buffer cache and inode cache of the kernel to cause all data to be written to the nbd-client.

\noindent{\bf \Unity\ File System: }
There can be cases where false sharing can occur for a user.  For example, she may be editing a document on her laptop, while listening to music on her phone.  If both the document and music file are stored on the same file system, this will result in many lease-holder switches if the file system is simply mounted on top of \UBD.  \UFS\ deals with the case by mapping each file onto its own DE.

\UFS\ is built by modifying the Minix $v3$ file system on Linux to maintain a mapping between file names and DE IDs and a mapping between disk blocks and offsets into those DEs.  On each I/O request, \UFS\ translates the request into a request for a DE and block ID and passes this information to the nbd-client.  We also modified the nbd-client to forward this information to the nbd-server.  Instead of using the same DE ID in all of its requests, the \UFS\ nbd-server then uses the proper DE ID and blockID in its request.  

The mapping between filenames, file system blocks and DE and block offsets is maintained in a special {\em translation} DE that stores these translations.  Again, different file system operations such as file creation, deletion or renames are all serialized through the translation DE, ensuring consistency of the \UFS\ name space.  However, because simultaneous access to different DEs can happen in arbitrary order, \UFS\ provides no consistency guarantees or restrictions on concurrent access to different files.  

\noindent{\bf Cloud node:}
The cloud node is a trivial client that never makes any requests and can never acquire the lease.  As a result, the cloud node always runs as a non-lease-holder and only the replication thread is active.

\section{Evaluation}
\label{sec:eval}

We evaluate three aspects of \Unity. First, we evaluate \Unity's use of bandwidth against NFS, NBD and Dropbox, a popular personal cloud storage service.  We also evaluate the amount of upload bandwidth usage of \Unity\ and compare this against its upload bandwidth if replication were done directly between devices.  Second, we measure \Unity's performance overhead versus the same alternative and show that \Unity\ gives comparable performance to other cloud and network storage options.  Finally, we evaluated the flexibility of DEs and how changing the mapping of DEs to storage abstractions can alleviate false sharing.  

\subsection{Evaluation Setup}

In our evaluation, we use four machines to represent a cloud node, a lease-holder user device, a non-battery powered device and a battery-powered mobile device in our lab environment. The cloud node and lease-holder user device are equipped with 3.4 GHz Intel i7-3770 CPUs and 16GB of memory. The non-battery powered user device is a machine with a 3.4 GHz Intel i7-2600 CPU and 16GB. Finally, the mobile device is a 2GHz AMD Dual Core Processor 3800+ with 2GB of RAM.  The coordinator is running on a virtual machine also in our lab environment.

The heartbeat period controls the frequency at which each of the devices sends and fetches state updates from the coordinator.  The heartbeat period is set to 10 seconds for the cloud node, 30 seconds for the non-battery user devices and 60 seconds for the mobile device.  We use a replication target of 3, which means that for each newly written block during a workload, one replica will exist on the lease-holder, another replica on the cloud node and the final replica will be on either the non-battery powered device or the mobile device. 

We evaluate \Unity\ with the following network configurations. One configuration uses an unconstrained gigabit network between the 4 nodes. We also used another configuration emulates a more realistic network setup where the non-battery nodes have high-end cable modem-like connections with 32 Mbps upload bandwidth and 128 Mbps download bandwidth.  The mobile node has a LTE-like connection with 16 Mbps download and 8 Mbps upload.  All network configurations are emulated with a D-Link DGS-1100 Gigabit EasySmart Switch, which supports the bandwidth control feature that can individually limit upstream and downstream bandwidth on each port.  


\subsection{Workloads}
We use three workloads in our evaluation.  To represent a compute-heavy workload with a mix of reads and writes, we compile the Emacs-24.2 editor.  The compilation tools are hosted on a local partition but the source files and target directory for the compilation are stored in \Unity.  For a streaming write workload we create a random 224MBytes file on a local partition and compress it using bzip2 onto the storage backed by \Unity.  Finally, for a streaming read workload, we read the compressed file from \Unity\ and decompress it using bzip2 back to a local file system.  All tests are done using the \UBD\ client, except for the evaluation of the flexibility of DEs where we compare the \UBD\ client with the \UFS\ client.


\subsection{Bandwidth utilization}




\begin{figure}[t]
  \centering
  \includegraphics[scale=.6]{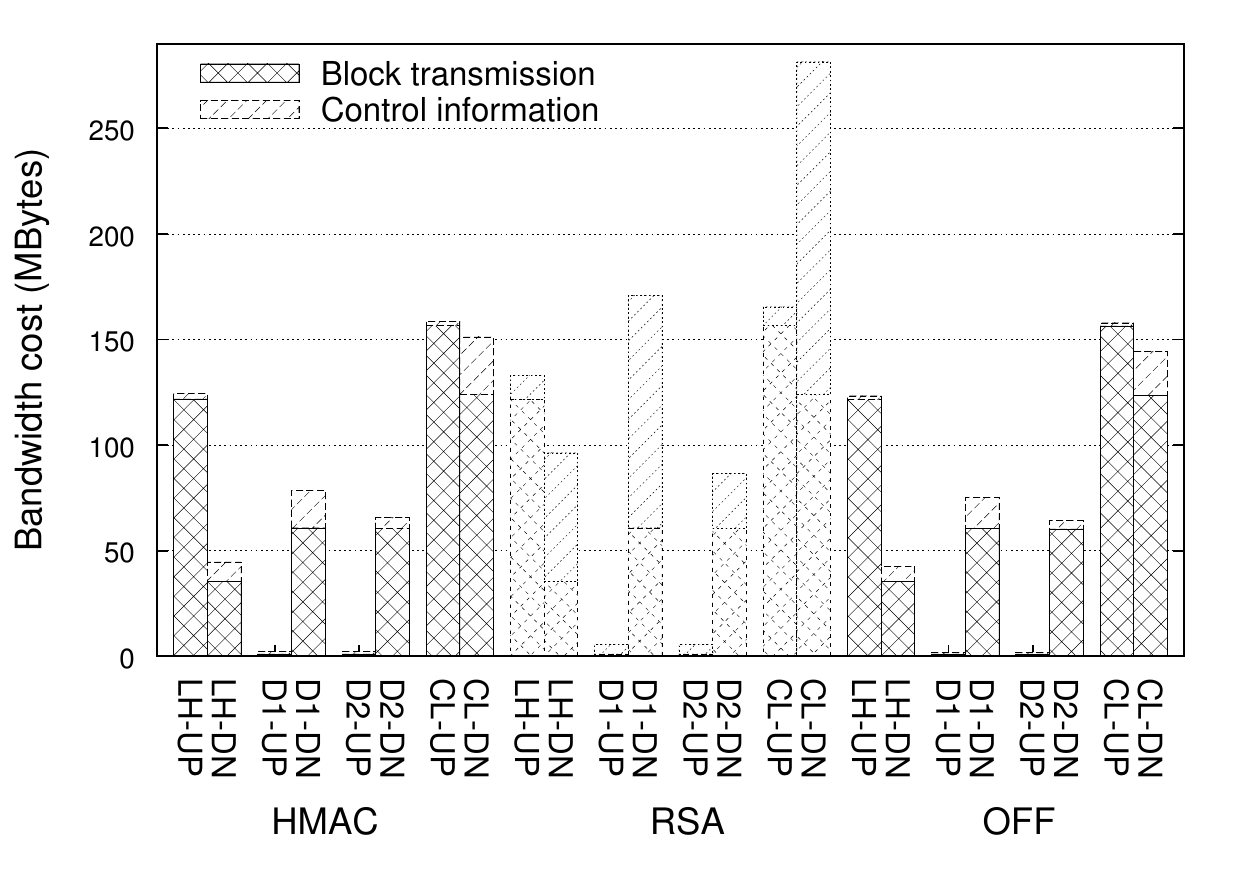}
  \vspace{-16pt}
  \caption{Compilation}
  \label{fig:compile}
  \vspace{-10pt}
\end{figure}
\begin{figure}[t]
  \centering
  \includegraphics[scale=.6]{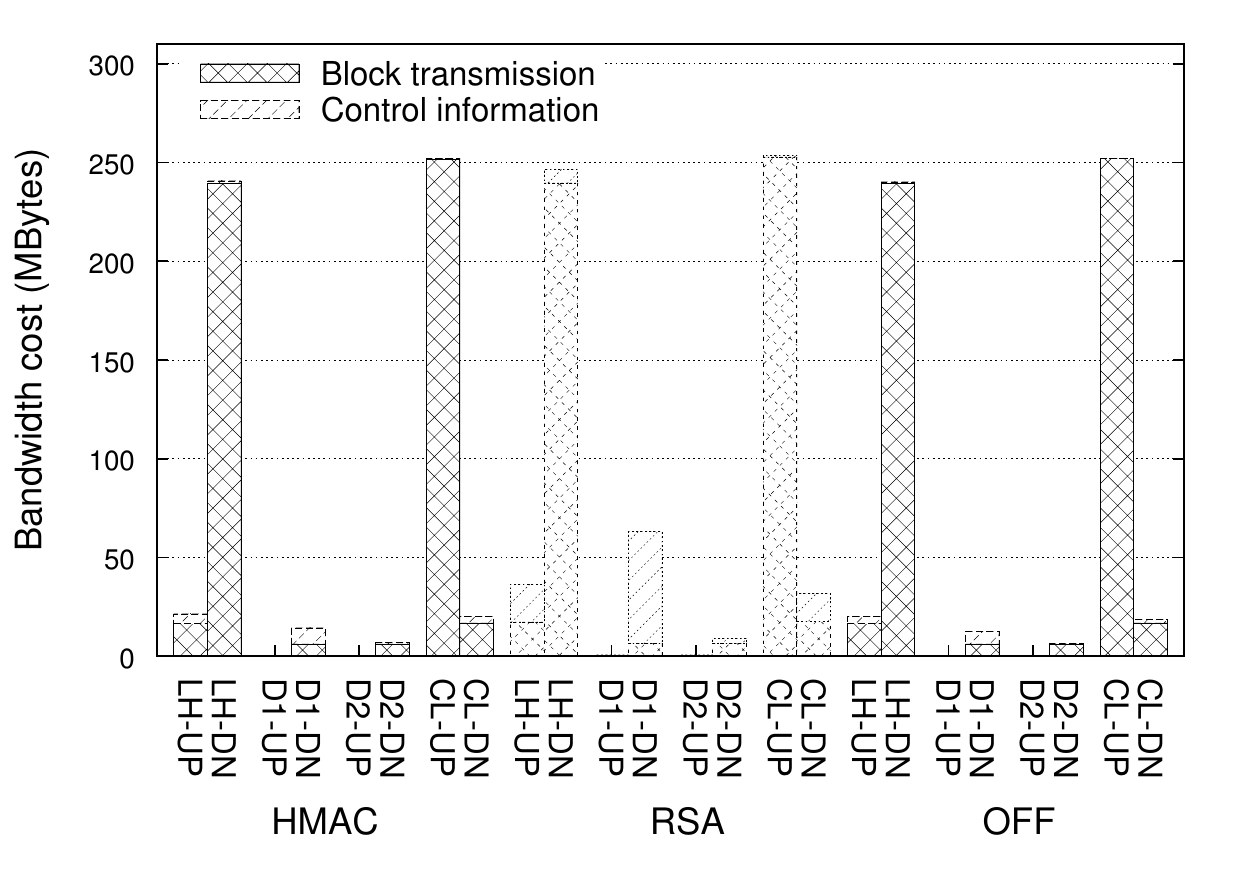}
  \vspace{-16pt}
  \caption{Streaming Read}
  \label{fig:sread}
    \vspace{-10pt}
\end{figure}
\begin{figure}[t]
  \centering
  \includegraphics[scale=.6]{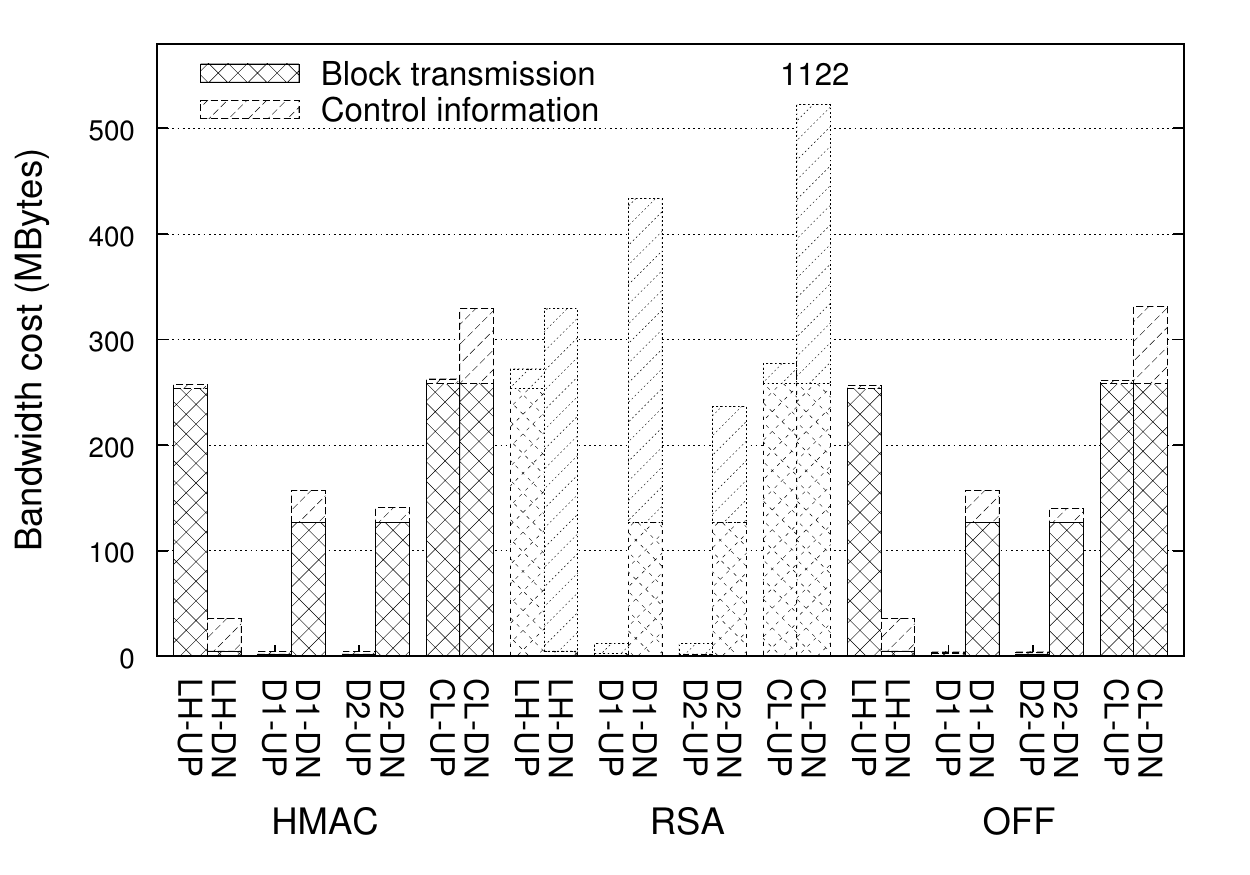}
  \vspace{-16pt}
  \caption{Streaming Write}
  \label{fig:swrite}
    \vspace{-10pt}
\end{figure}

We study how \Unity\ uses network bandwidth in each of our three workloads.  At the start of the workload, we assume all blocks have been replicated three times on the cloud node and the two non-lease-holder nodes. In addition, the lease-holder node executing the workload always fetches all needed blocks from the cloud node.  In addition, any block the lease-holder writes must be uploaded to the cloud node for the non-lease-holder nodes to replication.

We found that the size of the cryptographic signatures in our scheme incurs a significant amount of bandwidth utilization because the size of RSA signatures is relatively large compared to the size of the actual update sent to the coordinator.  Thus, we also modify \Unity\ to use HMAC-SHA1 (abbreviated to HMAC) which uses a 20 byte signatures, which is a lot lesser than the RSA's 256 byte signature.  While more efficient in terms of the bandwidth consumption, we note that the drawback of HMAC is that it is a symmetric signature scheme, meaning that all user devices must now be trusted since they all share a single signing and verification key.  We also compare these schemes to an implementation of \Unity\ where no digital signature scheme is used (OFF).  

Figure~\ref{fig:compile}, \ref{fig:sread} and~\ref{fig:swrite} show that the upload and download bandwidth consumption for each of the devices in all three workloads using each of the three signing schemes (LH represents the lease-holder, D1 is the non-battery powered devices, D2 is the mobile device and CL is the cloud node).  The graphs also show the bandwidth used for transferring block contents and control data in the form of state updates and lease-holder transfers.  In the compilation workload shown in Figure~\ref{fig:compile}, we see that the cloud node spends the upload bandwidth the most, followed by the lease-holder.  The lease-holder cannot avoid using upload bandwidth because it is writing new blocks due the compilation and must upload them to the cloud node.  However, the two non-lease-holder nodes use almost no upload bandwidth as they are just downloading the new blocks.  

We see that if RSA is used, the bandwidth used by control data dominates the workload because of the increase in the size of the update messages.  When we examined why the bandwidth usage was so high, we learned that our original decision allowing devices to specify which state updates they wanted to fetch from the coordinator by specifying a sequence number for a DE was problematic.  Because different non-lease-holder nodes replicate different blocks, they do not always replicate them in sequence. Sometimes, a block with a low sequence number might be remained as not fully replicated for some time. With this, devices may end up with repeatedly fetching the same update more than once.  A better approach might have been to maintain timestamps for each update and allow devices to only download updates no older than the configured time duration.

When comparing the control data bandwidth usage between the non-battery powered device (D1) and the mobile device (D2), we see the effect of the heartbeat interval on the download bandwidth consumption. Because D1 has a 30 second heartbeat period and D2 has a 60 second heartbeat period, we see that D1 spends significantly more download bandwidth on fetching updates than D2.  This effect is even more visible for the cloud node, which has a 10 second heartbeat period.

In Figure~\ref{fig:sread}, the streaming read workload shows much simpler behavior.  Since the workload has predominantly read heavy, we see that the lease-holder mainly downloads and the cloud node mainly uploads to send the blocks to the lease-holder.  Because very few blocks are newly written (mainly due to access time updates), D1 and D2 perform almost no replication and have very little upload or download bandwidth usage.

Finally, for the streaming write workload in Figure~\ref{fig:swrite}, we see that the usage of control bandwidth is even greater than other cases, especially for RSA.  We note that even for the large number of blocks newly written, the lease-holder spent very little upload bandwidth for control data compared to the amount of bandwidth spent to download data.  This again is an artifact caused by all nodes repeatedly downloading the same updates.

\begin{table*}[ht]
\caption{Bandwidth Consumption for \Unity, NFS NBD and Dropbox in MBytes (Down/Up)}
\vspace{0.1 in}
\centering
\begin{tabular}{| l | c | c | c | c | c | c |}
\hline
\multirow{2}{*}{Benchmark} &  \multicolumn{3}{| c |}{1G} & \multicolumn{3}{| c |}{Home} \\ \cline{2-7}
 & Compile & Sread & Swrite & Compile & Sread & Swrite \\ [1ex]
\hline\hline
NFS & 194 / 187  & 235 / 1 & 1 / 246 & 217 / 214 &  245 / 6 & 6 / 262 \\ \hline
NBD & 28 / 174 & 234 / 1 & 2 / 332 & 32 / 117 & 245 / 6 & 6 / 262 \\ \hline
Dropbox & 169 / 4 & 209 / 2 & 1 / 234 & 172 / 489 & 242 / 6 & 5 / 253 \\ \hline
\Unity & 80 / 50 & 263 / 32 & 349 / 269 & 167 / 52 & 265 / 36 & 682 / 284 \\ 
\hline
\end{tabular}
\label{table:compare_bandwidth_nfs_nbd_db}
\end{table*}

\noindent{\bf Comparison with NFS, NBD and Dropbox:}
We also compare the bandwidth consumption of \Unity\ with NFS, NBD and Dropbox.  For NFS and NBD, all data is placed on the server and only one client was used to run the workload.  In Dropbox, data is uploaded to the Dropbox server from one device. When the upload is completed, a second device where the workload would be run is connected to the Dropbox service so that it can start synchronizing.  Because Dropbox does not fetch block data on demand unlike NFS, NBD or \Unity, we wait for Dropbox to finish synchronizing data to the new device before starting the workload.  We enabled Dropbox's LAN Sync feature, which allows clients to transfer data directly to each other if they are on the same LAN.  \Unity\ runs with RSA signature scheme enabled.  

Table~\ref{table:compare_bandwidth_nfs_nbd_db} shows the bandwidth utilization measurement of the workloads across the different systems and network types.  We note that Dropbox has very low upload bandwidth usage on the compile workload under the unconstrained 1Gbps network. We think the reason is the repeated experimental run. We speculate that Dropbox likely detects that we are trying to upload identical files which are just uploaded recently and thus skip actual uploading.  \Unity\ is not as efficient as the other systems, largely due to the control data overhead.  However, in terms of upload bandwidth, it is more efficient than other systems, and it is only worse by a small amount for few cases.  Thus, we conclude that the price that \Unity\ pays for the added durability and security over single server or existing cloud systems is acceptable.

\noindent{\bf Energy Cost Saving:}
Uploading consumes more power for wireless devices than downloading. Thus, saving upload bandwidth consumption does not only improve performance, but also saves energy.  Using the data transfer energy model studied by Huang et. al.~\cite{DBLP:conf/mobisys/HuangQGMSS12}, we can estimate the energy saving offered by \Unity\ and compare it against that of having the lease-holder to replicate the block using direct peer-to-peer communication.  Our previous measurement shows that the block upload bandwidth usage of the compilation workload is about 122 MB. We consider three different network classes: wifi, 3G and LTE.  For the replication target set to three, the data should be uploaded at least twice. However, the design of \Unity\ allows it to be only uploaded once to the cloud node, because the third replicas can be downloaded by other device.  From this we can compute the energy saves.  We show the energy saving as the percentage of the whole battery capacity of a Nexus 4, which has a 3.8 V, 2100mAh battery, equivalent to 28,728 J of energy.


\begin{table}[ht]
\vspace{0.1 in}
\centering 
\begin{tabular}{l l l}
wifi & 552.75 J & 1.92 \%  \\
3G & 2,340.61 J & 8.15 \% \\
LTE & 1,129.68 J & 3.93 \% \\
\end{tabular}
\label{table:energy_cost_saving}
\caption{Energy savings offered by \Unity\ in J and as a percentage of a Nexus 4 battery.}
\end{table}

From this we can see that \Unity\ has the potential to save energy for wireless devices and the gain would proportionally increases as the replication target gets higher.


\begin{table*}[ht]
\caption{Performance Numbers for \Unity, NFS, NBD and Dropbox in Seconds}
\vspace{0.1 in}
\centering
\begin{tabular}{| l | c | c | c | c | c | c |}
\hline
\multirow{2}{*}{Benchmark} &  \multicolumn{3}{| c |}{1G} & \multicolumn{3}{| c |}{Home} \\ \cline{2-7}
 & Compile & Sread & Swrite & Compile & Sread & Swrite \\ [1ex]
\hline\hline
NFS & 209 & 31 & 31 & 349 & 65 & 34 \\ \hline
NBD & 95 & 33 & 32 & 124 & 80 & 54 \\ \hline
Dropbox & 1152 & 16 & 32 & 1155 & 128 & 32  \\ \hline
\Unity & 106 & 64 & 31 & 110 & 110 & 31  \\ 
\hline
\end{tabular}
\label{table:compare_perform_nfs_nbd_db}
\end{table*}

\subsection{Performance}


We record the run time overhead of \Unity\ when executing our workloads and compare it to that of NFS, NBD and Dropbox in Table~\ref{table:compare_perform_nfs_nbd_db}.  Since we have to wait for Dropbox to synchronize all files before starting the compilation, Dropbox's performance numbers are bad because the workload and file downloads are serialized, while they are interleaved on the other workloads.  This is the limitation of Dropbox because it does not take the file system activity into account when deciding what order to synchronize files in.  For the compilation workload, \Unity\ is generally competitive and is closest to NBD which \Unity\ is built based on.  For the streaming write benchmark, all operations happen locally on \Unity, NFS and Dropbox, and then are uploaded to the remote server or cloud node after completion, hence there is very little overhead.  Only NBD uploads blocks during the streaming write and suffers overhead.  In the streaming read workload, the main overhead on the fast 1Gbps network is due to the high cost of the RSA verification done for each block read. In addition, \Unity\ currently does not batch sequential block reads, thus requiring a round trip for each read.    However, this experiment demonstrates that our unoptimized \Unity\ prototype can still achieve performance comparable to that of NFS, NBD and Dropbox.

To evaluate how long \Unity\ takes to replicate data, we measure the average time it takes for a newly created block-version to reach the replication target.   With a constrained network, \Unity\ takes an average of 19.67 seconds over the 28,274 blocks written during the compilation workload and RSA signing.  The streaming write takes the average delay of 100.99 seconds to replicate the 60,811 blocks.  The increased delay of the streaming write results from the large number of blocks created.  Because the client thread is prioritized over the controller thread, no block replication is done until the workload completes.  Then, the sudden surge in network traffic as all devices try to replicate the large amount of data causes congestion, which slows down replication.

\subsection{DE Flexibility}

We measure DE flexibility by comparing between the performance using a single DE \Unity\ Block Device (\UBD) system and the performance running the same workload on \Unity\ file system (\UFS). As a workload, we compress the Linux kernel using gzip, while running \Unity\ on different folder with from 1 to 3 devices.  

\begin{figure}[!htb]
  \centering
  \includegraphics[scale=.6]{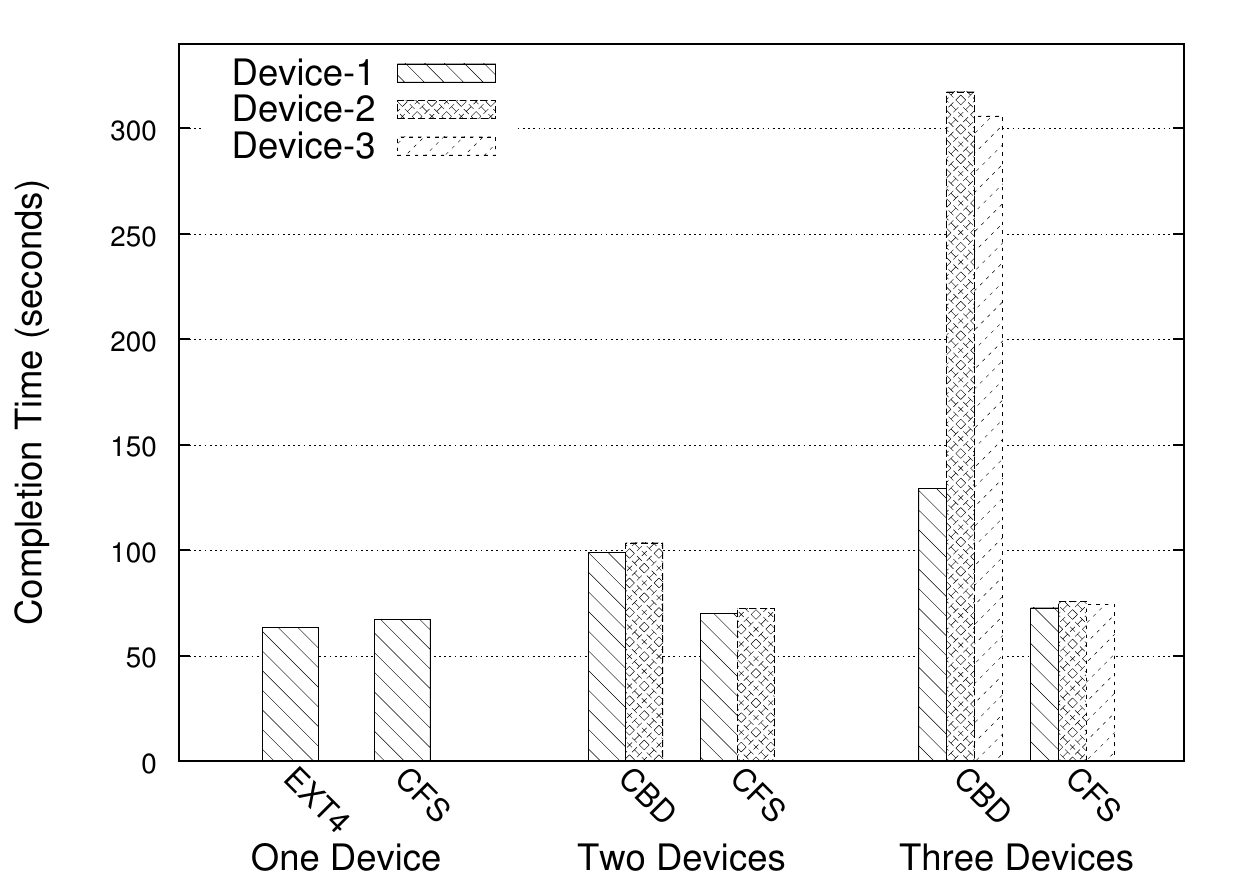}
  \caption{Compare Single DE with Caelus file system}
  \label{fig:fs}
\end{figure}

From Figure~\ref{fig:fs}, we can see that \UBD\ performs extremely poorly as the number of devices simultaneously accessing the DE increases.  In many cases, we observe that the contention is so bad that a device would barely get an I/O operation done in between lease-holder switches.  One device runs slightly faster than the others in the single DE case.  It is the device that the workload is started on, and it signals the other devices to start their workloads with a call to ssh.  As a result, the device likely runs exclusively on the DE for a while at starting phase and finished faster than the other devices.

In comparison, because \UFS\ maps each file to a unique DE, it eliminates the false sharing and virtually all lease-holder switches except for the specific DE storing mapping between DE IDs to actual files.  As a result, the performance of \UFS\ remains relatively unchanged with the number of devices.


We also compare \UFS\ with vanilla EXT4 on a single node to get the baseline overhead of our file system.  We can see that \UFS\ itself imposes very little overhead over a standard local file system.

\section{Related Work}
\label{sec:related}

There have been many existing works for securing distributed file systems in the presence of the untrusted server component such as SUNDR, Sirius, Plutus~\cite{sundr,sirius,plutus}. To protect data security, these systems are using cryptographic techniques such as digital signature, block checksum, encryption and decryption and so on in a similar way \Unity\ uses them. However, they cannot provide availability on the server failure. Several other works such as Farsite and OceanStore~\cite{farsite, oceanstore:pond} also deal with Byzantine failures. However, these Byzantine fault tolerant P2P file systems are using expensive quorum-based protocols flooding network with broadcast messages. Moreover, most of these distributed file systems assume multiple users, and therefore strong consistency is not feasible be realized. 

Hourglass, PDP, POR, DepSky and Hail~\cite{DBLP:conf/ccs/DijkJORST12,PDP,por,DBLP:conf/eurosys/BessaniCQAS11, hail} are exploring server side solution for untrusted cloud node issue either by applying cryptography techniques in a novel way or by using multiple cloud service providers. \Unity\ is trying to make use of the opportunity provided by the prevalent user devices. Also, Depot, Sporc and Venus~\cite{DBLP:conf/osdi/MahajanSLCADW10,DBLP:conf/osdi/FeldmanZFF10,DBLP:conf/ccs/ShraerCCKMS10} are very closely related to \Unity\ and try to provide similar Security and Durability guarantees as \Unity\ even with untrusted cloud. However, these systems provide consistency guarantees weaker than \Unity\ so that conflict can still occur and do not consider bandwidth efficiency much. Salus and Windows Azure~\cite{SalusNSDI13,DBLP:conf/sosp/CalderWONSMXSWSHUKEBMAAHHBDAMSMR11} provides a strong consistency guarantee which is similar to \Unity's. However, they are targeting for the enterprise environment. Therefore, they consider neither malicious cloud server nor bandwidth efficiency. 

\Unity's approach on security and durability guarantees against the untrusted cloud and the strong consistency model has been explored in our previous work~\cite{2012CCSW_Unity}. The work has been improved in its protocol and implementation to consider bandwidth efficiency, and more extensive evaluation studies have been added. Energy efficient storage systems on consumer electronic devices have been studied by projects like BlueFS~\cite{DBLP:conf/osdi/NightingaleF04}. \Unity\ is different from BlueFS by providing strong consistency guarantee through flexible DE abstraction and prefix semantic.

\section{Conclusion}
\label{sec:conclusion}

\Unity\ is a secure and durable personal cloud storage system that can minimize the bandwidth consumption of user's devices. \Unity\ allows users and cloud providers to establish a new relationship for user data management where the user devices provide durability and security guarantees and the cloud node provide high availability. By replicating blocks in a serialized manner, \Unity\ can provide strong consistency guarantee as well. Also, by making extensive use of Cloud's upload bandwidth instead of user devices', upload bandwidth consumption of user devices could be used much more efficiently. Our evaluation results show that \Unity\ can achieve security, durability and bandwidth efficiency at the small extra cost to exchange control information.

%
%

%

\bibliographystyle{abbrv}
\bibliography{bk}  
\end{document}